\documentclass[10pt,a4paper,final]{elsarticle}
\usepackage{mystyle}

\journal{Computational Materials Science}

\begin{document}

\begin{frontmatter}
\title{Phase Field Method for \\ Inhomogeneous Modulus Systems}

		\author[OSU]{Kamalnath Kadirvel \corref{firstauthor}}
	    \author[OSU,SJTU]{Pengyang Zhao}
		\author[OSU]{Yunzhi Wang \corref{correspondingauthor}}

\cortext[correspondingauthor]{Corresponding author, Email:wang.363@osu.edu}
\cortext[firstauthor]{Corresponding author, Email:kadirvel.1@buckeyemail.osu.edu}
\address[OSU]{Department of Materials Science and Engineering, The Ohio State University, 2041 College Road, Columbus, OH 43210, USA}
\address[SJTU]{Department of Engineering Mechanics, Shanghai Jiao Tong University, 800 Dongchuan Road, Shanghai, 200240, China}
\begin{abstract}
One of the advantages of the phase-field method (PFM) is its ability to incorporate elastic interactions that dominate solid-state processes including phase transformations and plastic deformation.
As mechanical equilibrium is attained much faster than chemical equilibrium, the former should be used as a constraint explicitly in deriving the governing equations of time-evolution of PFM order parameters.
Current models for elastically anisotropic and inhomogeneous media in the literature do not impose such a constraint in their governing equations.
In particular, they ignore the dependence of the total strain on the order parameters while evaluating the variational derivative of the elastic energy (VDEE). 
There is no mathematical proof to support this treatment and the fundamental thermodynamic consistency of such a models could be challenged. 
In this work,  we present a rigorous and physically transparent formulation of PFM governing equations for elastically anisotropic and inhomogeneous media in which the mechanical equilibrium is explicitly used as a constraint.
From our formulation, we show that VDEE computed in the  Leo-Lowengrub-Jou (1998) (LLJ) model is thermodynamically consistent.
We also show that the Wang-Jin-Khachaturyan (2003) (WJK) model has significant error in VDEE calculation.
We proposed a first-order correction term to the WJK model which nullifies the error in VDEE and makes the model thermodynamically consistent. 
Also, we computed the VDEE by a numerical method without using any assumptions and compared the numerical VDEE with the LLJ, WJK and modified WJK models.
\end{abstract}

\end{frontmatter}

%\linenumbers
	\section{Introduction}
%\begin{paragraph}{ }
 Phase-field method (PFM)  has emerged as an important tool to study microstructural evolution \cite{wang2010,steinbach2009,chen2002phase,moelans2008introduction} during various materials processes.
For solid-state processes such as solid-state phase transformations and plastic deformation, the elastic energy plays a crucial role in dominating the microstructural evolution. 
To deal with these solid-state processes in an elastically homogeneous medium, Khachaturyan derived a close-form solution of the elastic energy using the Green's function solution \cite{khachaturyan1969theory,khachaturyan1967,drkbook}.
This approach is known in the literature as microelasticity theory \cite{wang2010}.
In this approach, the elastic strain was written as a difference between the total strain and the inelastic strain.
Then the elastic energy is minimized instantaneously by relaxing the total strain for a given inelastic strain (or order parameters)   using the Green's function solution making the elastic energy a functional of only the order parameters.
As the elastic energy is directly written as a functional of the order parameters, mechanical equilibrium condition need not be solved explicitly, making this method computationally  efficient.
For an elastically inhomogeneous system, Y.U. Wang et al.~\cite{wang2002phase} introduced a "virtual strain"  field in the microelasticity model in order to map the stress field of the inhomogeneous system to a reference homogeneous modulus system, which then was solved using the Green's function solution.
 Shen et al.~\cite{shen2009} improved this model by developing a more efficient iterative approach to solve the virtual strain field. 
 %
 %\end{paragraph}

\longcomment{
Hence, it is important for PFM in order to compute accurately and efficiently the local stress variation within the microstructure.
Microelasticity theory \cite{drkbook} of phase-field is used to couple the microstructural evolution with the mechanical equilibrium.
As mechanical equilibrium is generally attained much faster than the chemical equilibrium, it should be used as a constraint for the phase-field dynamical equation.
}
 
%\begin{paragraph}{ }
Variational derivative of the elastic energy (VDEE) is required for phase-field governing equations such as Cahn-Hillard \cite{cahn1958}, Allen-Cahn \cite{allen1979} and Multi-Phase Field equations \cite{eiken2006}.
When the system is elastically homogeneous, i.e., the modulus is the same for all phases coexisting in the system, { VDEE can be derived as a function of order parameters } by following the procedure described by Khachaturyan \cite{drkbook}.
As elastic energy can be written as an explicit functional of order parameters only (i.e., the strain field is substituted using the Green’s function solution), it was possible to obtain exact closed form expression for VDEE. 
{
However, in the case of inhomogeneous modulus, the strain field can only be solved iteratively }\cite{wang2002phase,shen2009}  and thus the elastic energy cannot be written as a functional of only the order parameters. 
If mechanical equilibrium ($\nabla.\sigma=0$) is imposed as a constraint in the governing equations, the total strain field $\epsilon_{ij}(\vec{r})$ must be considered as a functional of the order parameters. 
Hence, many researchers have tried to obtain different approximations for VDEE in inhomogeneous modulus systems , which at least partially includes the effects of mechanical equilibrium.
For example,
% NEW PARAGRAPH
Onuki et al. \cite{onuki1991anomalously} assumed a weak dependence of the shear modulus on concentration and applied perturbation theory to perform a first-order expansion of the elastic free energy. 
His approximation was able to capture the effect of inhomogeneous modulus on spinodal decomposition qualitatively. 
Later, Leo-Lowengrub-Jou (referred to as LLJ) developed a phase-field model where the mechanical equilibrium ($\nabla.\sigma=0$) was solved accurately by a preconditioned conjugate gradient method \cite{leo1998}.
But the VDEE was derived from the elastic energy equation by assuming that the total strain field \Totalstrain is not a function of phase-field order parameters.
As mechanical equilibrium is attained much faster than chemical equilibrium, the strain field will respond instantaneously to any changes in the order parameter fields. 
Thus, the total strain field \Totalstrain must be considered as a function of the order parameters, as can be shown explicitly for the Eshelby inclusion problem as well as problems in elastically homogeneous media \cite{mura2013micromechanics}. 
Another model by Wang-Jin-Khachaturyan (referred to as WJK)  treats the elastic energy as a functional of the “virtual strain” field and order parameters, and the virtual strain is treated as a constant while deriving VDEE \cite{wang2003phase,wang2004phase}. 
These two distinct models (LLJ model \cite{leo1998} and WJK model \cite{wang2003phase,wang2004phase}) have different a priori assumptions and yield different values for VDEE although the calculated stress fields are identical.

In this article, we present, for the first time, a rigorous mathematical derivation for VDEE and evaluate quantitatively the difference between the two models in the literature. We also computed the VDEE numerically and compared with the analytical expressions in the literature.
%\end{paragraph}

\longcomment{
We need a closed form expression of the variational derivative $\veq{E^{el}}{\theta}$ under the constraint $\nabla. \sigma=0$ for evolving the microstructure $\theta(\vec{r})$ \cite{\rf{14}}.
In the physical systems where the modulus of the precipitate is very close to that of the matrix, the entire system can be approximated as elastically homogeneous where the variational derivative $\veq{E^{el}}{\theta}$ can be derived in a closed-form expression \cite{drkbook}. 

This assumption is reasonable for many systems under its operating conditions.
For example, in nickel-based alloys the modulus difference between the precipitate and matrix is only about 15\% \cite{\rf{46}}.
% \toref{PhilMagNing,Homo-modulus} \cite{\rf{6}}
But in the studies were the system is elastically inhomogeneous \cite{\rf{21},\rf{13}} , in the evaluation of $\veq{E^{el}}{\theta}$, the mechanical equilibrium was not imposed as constraint explicitly.
Many researchers have used the Leo’s (see, e.g., \cite{hu2001,steinbach2006multi,guru2007,ammar2009combining,gaubert2010coupling}) and Zhou’s (see, e.g., \cite{wang2003phase,wang2004effects,wang2004phase,zhou2009gamma}) approximations to microstructural evolutions in elastically inhomogeneous systems. 
It should be noted that Steinbach et al. \cite{steinbach2006multi} mentions about imposing the mechanical equilibrium ($\sigma^\alpha=\sigma^\beta$) at interfaces between $\alpha$ and $\beta$ phases in their multi-phase field model, but the mechanical equilibrium constraint is imposed after performing VDEE. 
}

	\section{Method}
	    \subsection{Background}
%\begin{paragraph}{ }
In PFM, the total free energy is formulated as a functional of the so-called order parameter fields.
For simplicity, let us consider a binary alloy system with a concentration field $c(\vec{r})$ and a structural order parameter field $\eta(\vec{r})$ as the order parameters.
The total free energy can be decomposed into chemical (including bulk and interfacial contributions) and elastic free energies:
\
\begin{align}
F[c(\vec{r}),\eta(\vec{r})]&=F^\text{chem}[c(\vec{r}),\eta(\vec{r})]+E^\text{el}[c(\vec{r}),\eta(\vec{r})] \\
		E^\text{el}[c(\vec{r}),\eta(\vec{r})] &=\int \dfrac{1}{2} C_{ijkl}(c(\vec{r}),\eta(\vec{r})) \epsilon_{ij}^{el}(\vec{r}) \epsilon_{kl}^{el}(\vec{r}) dV 
\end{align}
where $C_{ijkl}$ is the elastic modulus and $\epsilon^\text{el}_{ij}$ is the elastic strain of the system.
The order parameters $c(\vec{r})$ and $\eta(\vec{r})$ are evolved through Cahn-Hilliard \cite{cahn1958} and Allen-Cahn \cite{allen1979} equation, respectively.
\begin{align}
\deq{c}{t} 
&= \nabla \left( M \nabla \left( 
\veq{F^\text{chem}}{c} + \veq{E^\text{el}}{c} \right) \right)\\
\deq{\eta}{t}
&= - L \left( \veq{F^\text{chem}}{\eta} + \veq{E^\text{el}}{\eta} \right)
\end{align}
where $M$ is the chemical mobility and $L$ is the kinetic coefficient for structural evolution.
In general, the elastic energy $E^\text{el}$ is a functional of $c(\vec{r})$ or $\eta(\vec{r})$ or both, depending on the coupling between the eigenstrain and order parameters.
In this work, we consider that the elastic energy is only a functional of the structural order parameter field $\eta$ for simplicity.
In case of coupling with structural order parameters, typically we use the interpolation function $\theta(\vec{r})=\eta^3(6\eta^2-15\eta+10)$ \cite{zhou2010} to couple the transformation strain with the order parameter.
By using the chain rule, we can write 
$$\veq{E^\text{el}}{\eta} = \deq{\theta}{\eta} \veq{E^\text{el}}{\theta} .$$
\longcomment{
The coupling between $\theta(\vec{r})$ and $\eta(\vec{r})$ can be originated from the underlying transformation pathways \cite{zhao2017}.
}
In evaluating the variational derivative \VDEE, mechanical equilibrium should be imposed as a constraint.
In other words, infinitesimal changes in the elastic strain $\delta \epsilon^\text{el}_{ij}(\vec{r'})/\delta \theta(\vec{r})$ imposed by mechanical equilibrium must be considered while evaluating the VDEE $\delta E^\text{el}/\delta \theta(\vec{r})$.
For the remainder of the paper, we will focus only on the evaluation of the VDEE  with analytical and numerical models.
%
%\end{paragraph}

	    \subsubsection{VDEE for homogeneous modulus systems}

%\begin{paragraph}{ }
%
When the modulus of the system is homogeneous, mechanical equilibrium can be solved using Green's function solution.
Let us consider the modulus of the system to be $C^o_{ijkl}$.
The misfit strain between the matrix phase ($\eta=0$) and the precipitate phase ($\eta=1$) is $\epsilon^{oo}_{ij}$, also called eigenstrain.
\longcomment{ Eigenstress is defined as $\sigma^{oo}_{ij}=C^o_{ijkl}\epsilon^{oo}_{kl}$. }
The transformation strain field is defined as $\epsilon^T_{ij}(\vec{r})=\epsilon^{oo}_{ij} \theta(\vec{r})$.
\longcomment{ Transformation stress is defined as $\sigma_{ij}^T(\vec{r})=C^o_{ijkl} \epsilon_{ij}^T (\vec{r})$. }
The elastic strain $\epsilon^\text{el}_{ij}(\vec{r})$ is defined as the difference between the total strain $\epsilon_{ij}(\vec{r})$ and the transformation strain i.e., $\epsilon^\text{el}_{ij}(\vec{r}) = \epsilon_{ij}(\vec{r}) - \epsilon^T_{ij}(\vec{r}) $.
Then the stress field is given by $\sigma=C^o_{ijkl} \epsilon^\text{el}_{kl}$.
The mechanical equilibrium equation is given by
\begin{align}
\nabla \cdot \boldsymbol{\sigma} = 0 ; \qquad \sigma_{ij}=C^o_{ijkl}(\epsilon_{ij}-\epsilon^T_{ij})    
\end{align}
By using Green's function approach, the total strain field can be derived as \cite{drkbook},
\begin{align}
\label{stn_theta}
			\epsilon_{ij}(\vec{r}) = \bar{\epsilon}_{ij} 
			+\frac{1}{2} \int \dfrac{d^3k}{(2\pi)^3} e^{i \vec{k}.\vec{r}}
			[G_{ijkl}+G_{jikl}] \tilde{\sigma}^T_{kl}  
\end{align}
where  $\bar{\epsilon}_{ij}$ is the average total strain, 
$\tilde{\sigma}_{kl}^T(\vec{k})= \int \sigma_{kl}^T(\vec{r}) e^{-i\vec{k}.\vec{r}} d\vec{r} $ is the Fourier transform of $\sigma_{ij}^T(\vec{r})=C^o_{ijkl} \epsilon_{kl}^T(\vec{r})$,
the Green's operator $G_{ijkl}=n_i \Omega_{jk} n_l$, 
$\Omega_{jk}^{-1}=n_i C^o_{ijkl} n_l$ and $\hat{n}$ is a unit vector along $\vec{k}$. Elastic energy of the system is given by,
\begin{align}
 E^\text{el} 
  = \int  
\frac{1}{2} C^o_{ijkl} (\epsilon_{ij}-\epsilon_{ij}^T) (\epsilon_{kl}-\epsilon_{kl}^T) 
 dV    
\end{align}
Substituting Eq.\eqref{stn_theta} in the above equation and simplifying the expression gives us the elastic energy as a functional of $\theta(\vec{r})$ \cite{drkbook}.
\begin{align}
\label{Eel_theta}
E^\text{el}[\theta(\vec{r})] = 
    \frac{1}{2} \int  B(\hat{n}) \tilde{\theta}(\vec{k}) \tilde{\theta}^*(\vec{k}) \din{k}
\end{align}
where  $\tilde{\theta}(\vec{k})= \int \theta(\vec{r}) e^{-i\vec{k}.\vec{r}} d\vec{r}$  and $B(\hat{n})$ is defined as	
\longcomment{ Need to write the expression of Fourier transform}	
\begin{align}
			B(\hat{n})=\begin{cases}
				0 & \text{-if } \hat{n}=0 \\
				C^o_{ijkl}\epsilon_{ij}^{oo}\epsilon_{kl}^{oo}- G_{ijkl}(\hat{n}) \sigma_{ij}^{oo} \sigma_{kl}^{oo} 
				& \text{-if } \hat{n}\neq 0\\
			\end{cases}
\end{align}
where $\sigma_{ij}^{oo}=C^o_{ijkl}\epsilon_{kl}$.
Here, we assume a stress-free boundary condition.
As the elastic energy $E^\text{el}[\theta(\vec{r})]$ is expressed as an explicit functional of the order parameter, its variational derivative can be evaluated readily and is given by
\begin{align}
\label{var_der_homo}
\veq{E^\text{el}}{\theta(\vec{r})} = 
-\sigma^\text{app}_{ij} \epsilon^{oo}_{ij} 
+ \int \din{k} e^{i \vec{k}.\vec{r}} 
[ B(\vec{n}) \tilde{\theta}(\vec{k}) ].
\end{align}
%
%\end{paragraph}

	    \subsubsection{ Mechanical equilibrium solver for inhomogeneous modulus systems}
\label{yao_shen_sec}

%\begin{paragraph}{ }
The total strain field $\epsilon_{ij}(\vec{r})$ of an inhomogeneous modulus system has to be solved in an iterative manner.
We will briefly describe the solver used in our work to calculate local stress and local strain fields under mechanical equilibrium \cite{shen2009}.
Let us write the local variation in the modulus $C_{ijkl}$ as 
$$C_{ijkl}(\vec{r})=C^o_{ijkl}-\Delta C_{ijkl}(\vec{r}).$$
Assume a virtual system, whose homogenous modulus is $C^o_{ijkl}$ and have a tranformation strain \vrs .
Let us call it "virtual strain" in order to distinguish from the transformation strain in the real system.
Both the stress field $\sigma_{ij}(\vec{r})$ and the displacement field $u_{i}(\vec{r})$ of the virtual system are assumed to be identical to the real inhomogeneous system.
As the displacement field is identical, the total strain field $\epsilon_{ij}(\vec{r})$ of the reference system is also identical to the real system.
The virtual strain $\epsilon_{ij}^o(\vec{r})$ is solved iteratively as follows: 
Initially we can assume that $(\epsilon^o_{ij})^0=\epsilon_{ij}^T$ and for every $p^{th}$-iteration, the virtual strain is updated as
\begin{align}
\nonumber
(\epsilon^o_{ij})^{(p+1)} = &(\epsilon^o_{ij})^{(p)} + L S^o_{ijkl} \Delta C_{klmn}
\left[
\frac{1}{2} \int \frac{d^3k}{(2\pi)^3}  
[G_{mnpq}+G_{nmpq}] C^o_{pqrs} (\tilde{\epsilon}_{rs}^o)^{(p)} e^{i\vec{k}.\vec{r}} \right] \\
\label{vrs_iter}
& +L\left[S^o_{ijkl} \Delta C_{klmn}( \bar{\epsilon}_{mn} - \epsilon^T_{mn} ) 
-(\epsilon^o_{ij})^{(p)}+\epsilon^T_{ij} \right]
\end{align}
where $L$ is the step-length.
Once the virtual strain is converged, the total strain field can be solved using 
\begin{align}
\label{tls_inhom}
\epsilon_{ij}=\bar{\epsilon}_{ij} +
\frac{1}{2} \int \frac{d^3 k}{(2\pi)^3}
[G_{mnpq}+G_{nmpq}] C^o_{pqrs} \tilde{\epsilon}_{rs}^o e^{i\vec{k}.\vec{r}}
\end{align}
and subsequently the stress field $\sigma_{ij}(\vec{r})$ is computed using
%\begin{align}
$\sigma_{ij}=C^o_{ijkl}(\epsilon_{kl}-\epsilon_{kl}^o)$.    
%\end{align}
\longcomment{
Alternately, the stress field can also be calculated directly on the real elastically inhomogeneous system, i.e., $\sigma_{ij}=C_{ijkl}(\vec{r}) (\epsilon_{kl}-\epsilon_{kl}^{T})$. This equation can be rearranged into 
\begin{align}
 \epsilon_{ij}=\epsilon^T_{ij} + S_{ijkl}(\vec{r})\sigma   
\end{align}
where $S_{ijkl}(\vec{r}) = ( C_{ijkl}(\vec{r}) )^{-1}$ is the compliance of the system.
}
%\end{paragraph}
%
%\begin{paragraph}{ }
%
In this case, there is no closed-form analytical expression for the strain field and hence one cannot compute the variational derivative analogous to the previous subsection.
%\end{paragraph}

	    \subsection{VDEE for inhomogeneous modulus systems}
\label{anal_expressions}

%\begin{paragraph}{ }
 Many studies have used the elastically inhomogeneous systems for simulating microstructural evolutions (see, e.g., \cite{leo1998,zhou2008,steinbach2006multi,ammar2009combining,wang2004effects,wang2004phase,guru2007,hu2001,wang2003phase,ratz2006surface,kadirvel2021phase,kadirvel2023microstructural,kloenne2021high}).
The elasticity models in all these kind of works can be categorised into two kinds:
 (i) LLJ model:(see, e.g., \cite{leo1998,hu2001,steinbach2006multi,guru2007,ammar2009combining,gaubert2010coupling}) . 
 (ii) WJK model: \cite{wang2003phase,wang2004effects,wang2004phase,ratz2006surface,zhou2009gamma,boyne2011concurrent,boyne2011surface}.
%New paragraph
The LLJ model \cite{leo1998} ignored the dependence of the total strain field $\epsilon_{ij}(\vec{r})$ on the order parameter \OrderParameter while evaluating the VDEE (\VDEE).
In other words, the total strain field $\epsilon_{ij}(\vec{r})$ and the order parameter field \OrderParameter were treated as independent variables while evaluating the variational derivative.
Infinitesimal changes in the order parameter field ($\delta\theta(\vec{r})$)  would cause infinitesimal changes in the total strain field ($\delta\epsilon_{ij}(\vec{r})$) due to mechanical equilibrium and this effect should be included while evaluating the variational derivative.
For instance, in an elastically homogeneous system, if we calculated the variational derivative of \Totalstrain  with respect to \OrderParameter , using Eq.\eqref{stn_theta}, we would get
\begin{align}
    \veq{\epsilon_{ij}(\vec{r'})}{\theta(\vec{r})}= \int \din{k} \frac{G_{ijkl}+G_{jikl}}{2} C^o_{klmn} \epsilon_{mn}^{oo} e^{-i\vec{k}.(\vec{r}-\vec{r'})}.
\end{align}
This clearly demonstrates that the infinitesimal changes in the order parameter \OrderParameter and in the total strain \Totalstrain are related under mechanical equilibrium for elastically homogeneous systems.
Similar relation may exist for elastically inhomogeneous systems, but as we do not have a closed form expression for the total strain, we can not evaluate it analytically.
Hence, it is not clear a priori that neglecting such a relationship will still give a reasonable expression for the variational derivative of the elastic energy (VDEE). 
 The mechanical equilibrium (\MechanicalEquilibrium) was only used as a kinetic equation to be evolved alongside the order parameter $\theta(\vec{r})$ in the LLJ model.
But the computed strain field \Totalstrain was substituted into VDEE expression and used in the evolution of order parameters.
%\end{paragraph}

%\begin{paragraph}{ }
In the WJK model \cite{wang2003phase}, the VDEE was computed in terms of the virtual strain.
As we have a closed form expression of \Totalstrain in terms of the virtual strain $\epsilon^o_{ij}(\vec{r})$, we can compute the variational derivative $\delta \epsilon_{ij} (\vec{r'})/\delta \epsilon_{mn}^o (\vec{r})$.
By assuming the elastic energy to be a functional of the virtual strain \vrs and neglecting the infinitesimal changes  $\delta\epsilon^o_{ij}(\vec{r})$ due to infinitesimal changes in \OrderParameter, variational derivative of the elastic energy was evaluated.
At each time step, the virtual strain \vrs  was solved using the numerical iteration method described by Eq.\eqref{vrs_iter} and it was used as a constant for calculating the VDEE for that time moment.
The VDEE can be regarded as the $0^\text{th}$ order expansion of the variational derivative in terms of the virtual strain in the WJK model.
%
%\end{paragraph}

%\begin{paragraph}{ }
%The above two models (LLJ and WJK) give different values for the VDEE as will be shown later.
%
%\end{paragraph}
 Let us consider the modulus to be a function of the order parameter, i.e., $C_{ijkl}(\theta)= C^o_{ijkl} - \Delta C_{ijkl}(\theta)$, where $C^o_{ijkl}$ is constant.
 In LLJ model \cite{leo1998}, the VDEE is derived as 
 \begin{align}
     \label{veq_leo}
\veq{E^{el}}{\theta(\vec{r})} = - \frac{1}{2} \deq{\Delta C_{ijkl}}{\theta} (\epsilon_{ij}-\epsilon_{ij}^{oo}\theta) (\epsilon_{kl}-\epsilon_{kl}^{oo}\theta) - \sigma_{ij} \epsilon_{ij}^{oo}
 \end{align}
 Note that the above expression is the same as that given in Eq.(36) of Ref.\cite{guru2007}.
In WJK model \cite{wang2003phase}, the VDEE becomes
\begin{align}
\label{veq_zhou}
\veq{E^{el}}{\theta(\vec{r})}=\frac{1}{2} C^o_{ijmn} \deq{\Delta S_{mnpq}}{\theta} C^o_{ijkl} [\epsilon^o_{ij}-\epsilon_{ij}^T][\epsilon^o_{kl}-\epsilon_{kl}^T] 
- \frac{1}{2} (C^o_{ijmn} \Delta S_{mnpq} C^o_{ijkl} -C^o_{ijkl}) [\epsilon_{kl}^o-\epsilon_{kl}^T]\deq{\epsilon^T_{ij}}{\theta}.
\end{align}
where $\Delta S_{ijkl}= (\Delta C_{ijkl})^{-1}$.
Note that the above expression is the same as that given in Eq.(12) of Ref.\cite{zhou2008}. 
The virtual strain \vrs was solved using the numerical iteration (Section~\ref{yao_shen_sec}) and treated as constant for the given time step ($0^\text{th}$ order).
Let us refer to this expression as "WJK model".
As will be shown later, this assumption introduces considerable amount of error in the calculation.
This expression can be improved easily by assuming $\deq{\epsilon_{ij}^o}{\theta}=\deq{\epsilon_{ij}^T}{\theta}$.
Then the VDEE becomes
\begin{align}
\nonumber
\veq{E^{el}}{\theta(\vec{r})} 
&= \frac{1}{2} C^o_{ijmn} \deq{\Delta S_{mnpq}}{\theta} C^o_{ijkl} [\epsilon^o_{ij}-\epsilon_{ij}^T][\epsilon^o_{kl}-\epsilon_{kl}^T] \\
\label{veq_Czhou}
&-\deq{\epsilon^T_{ij}}{\theta} \left( 
\int \din{k}
C^o_{ijmn} G_{mnkl} C^o_{klts} \tilde{\epsilon}_{ts}^o 
e^{i\vec{k}.\vec{r}}
- C^o_{ijkl}(\epsilon_{kl}^o-\bar{\epsilon}_{kl}^o)
\right)
\end{align}
The above expression from now on will be referred to as "modified WJK".
%\end{paragraph}

\longcomment{
In terms of the equations,
 \begin{align*}
 E^{el}[\theta(\vec{r}), \{ \epsilon_{ij}(\vec{r}) \}] 
 &= \int \underbrace{ 
\frac{1}{2} C_{ijkl}(\theta) (\epsilon_{ij}-\epsilon_{ij}^{oo}\theta) (\epsilon_{kl}-\epsilon_{kl}^{oo}\theta) 
}_{f} dV \\
 \veq{E^{el}}{\theta} &= \deq{f}{\theta} - \underbrace{\nabla.\left( \deq{f}{\nabla \theta} \right) }_{zero} \\
 \nonumber
 &= \frac{1}{2} \deq{C_{ijkl}}{\theta} (\epsilon_{ij}-\epsilon_{ij}^{oo}\theta) (\epsilon_{kl}-\epsilon_{kl}^{oo}\theta) +
 \frac{1}{2} C_{ijkl} (\deq{\epsilon_{ij}}{\theta}-\epsilon_{ij}^{oo}) (\epsilon_{kl}-\epsilon_{kl}^{oo}\theta)  \\
 & \qquad \qquad 
 +\frac{1}{2} C_{ijkl} (\epsilon_{ij}-\epsilon_{ij}^{oo}\theta) (\deq{\epsilon_{kl}}{\theta}-\epsilon_{kl}^{oo}) \\
 \intertext{Using $\deq{\epsilon_{ij}}{\theta}=0$ and combining $2^{nd}$ and $3^{rd}$ term}
 &=\frac{1}{2} \deq{C_{ijkl}}{\theta} (\epsilon_{ij}-\epsilon_{ij}^{oo}\theta) (\epsilon_{kl}-\epsilon_{kl}^{oo}\theta) +  C_{ijkl} (\epsilon_{ij}-\epsilon_{ij}^{oo}\theta) (-\epsilon_{kl}^{oo})
 \end{align*}
 Using $\sigma_{kl}=C_{ijkl}(\epsilon_{ij}-\epsilon_{ij}\theta)$
 \begin{align}
 \label{varEel}
 \veq{E^{el}}{\theta} =\frac{1}{2} \deq{C_{ijkl}}{\theta} (\epsilon_{ij}-\epsilon_{ij}^{oo}\theta) (\epsilon_{kl}-\epsilon_{kl}^{oo}\theta) - \sigma_{ij} \epsilon_{ij}^{oo}
 \end{align}
}% Comment region

		\section{Physically transparent derivation of the VDEE }
	\label{physical_derivation}

	%\begin{paragraph}{ }
	We formulate the elastic energy $E^{el}$ as a functional of $C_{ijkl}$ and the elastic strain $\epsilon_{ij}^{el}$, which are in turn a functional of the order parameter \OrderParameter.
	
	\begin{align}
	E^{el}[C_{ijkl}[\theta(\vec{r})],\epsilon_{ij}^{el}[\theta(\vec{r})]]=\int \dfrac{1}{2} C_{ijkl} \epsilon_{ij}^{el} \epsilon_{kl}^{el} dV \\
	\intertext{Its variational derivative is given by}
	\label{varEl2}
	\veq{E^{el}}{\theta(\vec{r})} 
	= \int \veq{E^{el}}{C_{ijkl}(\vec{r'})}  \veq{C_{ijkl}(\vec{r'})}{\theta(\vec{r})} d^3r' +
	\int \veq{E^{el}}{\epsilon_{ij}^{el}(\vec{r'})} \veq{\epsilon_{ij}^{el}(\vec{r'})}{\theta(\vec{r})} d^3r' .
	\end{align}
	Evaluating the first term using $C_{ijkl}(\vec{r'})=C^o_{ijkl} - \Delta C_{ijkl} (\theta(\vec{r'}) )$, we have
	\begin{align}
	\nonumber
	\int \veq{E^{el}}{C_{ijkl}(\vec{r'})}  \veq{C_{ijkl}(\vec{r'})}{\theta(\vec{r})} d^3r' 
	&= \int \left( 
	\frac{1}{2} \epsilon_{ij}^{el}(\vec{r'}) \epsilon_{kl}^{el}(\vec{r'})
	\right)
	\left( 
	-\deq{ \Delta C_{ijkl} }{\theta(\vec{r'})} \delta(\vec{r'}-\vec{r})
	\right) d^3r'\\
	\label{deq6}
	&= - \frac{1}{2} \deq{ \Delta C_{ijkl} }{\theta} \epsilon_{ij}^{el}(\vec{r}) \epsilon_{kl}^{el}(\vec{r})
	\end{align}
	Evaluating the second term in Eq.\eqref{varEl2}, with $\epsilon_{ij}^{el}=\epsilon_{ij}-\epsilon_{ij}^{oo} \theta$ and
	$ \veq{E^{el}}{\epsilon_{ij}^{el}(\vec{r'})} = \sigma_{ij} (\vec{r'})$, we have  
	\begin{align}
	\nonumber
	\int \veq{E^{el}}{\epsilon_{ij}^{el}(\vec{r'})} \veq{\epsilon_{ij}^{el}(\vec{r'})}{\theta(\vec{r})} d^3r' &= 
				\int \sigma_{ij}(\vec{r'}) \left[ \veq{\epsilon_{ij}(\vec{r'})}{\theta(\vec{r})} - \epsilon_{ij}^{oo} \delta(\vec{r'}-\vec{r}) \right] d^3r' \\
				\label{deq5}
			&= \int \sigma_{ij}(\vec{r'}) \veq{\epsilon_{ij}(\vec{r'})}{\theta(\vec{r})} d^3r' - \sigma_{ij} \epsilon_{ij}^{oo}
	\end{align}
The first term in the above equation is zero under mechanical equilibrium as shown in the subsequent equations.
By using the chain rule we can write,
\begin{align}
    \veq{\epsilon_{ij}(r')}{\theta(r)}
	= \int \veq{\epsilon_{ij}(r')}{\epsilon^o_{mn}(r'')} \veq{\epsilon^o_{mn}(r'')}{\theta(r)} d^3r''
\end{align}
As the total strain is an explicit functional of the virtual strain Eq.\eqref{tls_inhom}, we can compute $\veq{\epsilon_{ij}(r')}{\epsilon^o_{mn}(r'')}$ explicitly. 
However, as the virtual strain is an implicit functional of $\theta(\vec{r})$, we can't compute $\veq{\epsilon^o_{mn}(r'')}{\theta(r)}$ explicitly.
Simplifying the first term of the equation Eq.\eqref{deq5} using the chain rule,
\begin{align}
	\label{deq4}
\int \sigma_{ij}(\vec{r'}) \veq{\epsilon_{ij}(\vec{r'})}{\theta(\vec{r})} d^3r' 
&=  \int \left[
\int \sigma_{ij}(\vec{r'}) \veq{\epsilon_{ij}(r')}{\epsilon^o_{mn}(r'')} d^3r'
\right]
\veq{\epsilon^o_{mn}(r'')}{\theta(r)}  d^3r'' 
\end{align}
As mentioned in Section~\ref{yao_shen_sec}, the total strain in terms of the virtual strain is given by
\begin{align*}
\epsilon_{ij}&=\bar{\epsilon}_{ij}+  \int \din{k} \frac{[G_{ijkl}+G_{jikl}]}{2} C^o_{klmn} \tilde{\epsilon}^o_{mn} e^{i\vec{k}.\vec{r}}.
\end{align*}
By taking the variational derivative of the total strain w.r.t the virtual strain, we get
\begin{align}
	\veq{\epsilon_{ij}(\vec{r'})}{\epsilon_{mn}^o(\vec{r})}= \int \din{k} H_{ijkl} C^o_{klmn} e^{-i\vec{k}.(\vec{r}-\vec{r'})}
\end{align}
where  $H_{ijkl}=1/2(G_{ijkl}+G_{jikl})$.
Using the above equation we can write
\begin{align}
	\nonumber
	\int \sigma_{ij}(\vec{r'}) \veq{\epsilon_{ij}(r')}{\epsilon^o_{mn}(r)} d^3r' 
	&= \iint \sigma_{ij}(\vec{r'})  H_{ijkl} C^o_{klmn} e^{-i\vec{k}.\vec{r}}  e^{i\vec{k}.\vec{r'}} \din{k} d^3r' \\
	\nonumber
	&= \int \left[ \int \sigma_{ij}(\vec{r'}) e^{i \vec{k}.\vec{r'} } d^3r' \right] H_{ijkl} C^o_{klmn} e^{-i\vec{k}.\vec{r}} \din{k} \\
	\label{deq7}
	&= \int \tilde{\sigma}_{ij} H_{ijkl} C^o_{klmn} e^{-i\vec{k}.\vec{r}} \din{k}
\intertext{As i and j are dummy indices, substituting for $H_{ijkl}$ and swapping the indices in the RHS we have,}
	\label{deq8}
	\int \sigma_{ij}(\vec{r'}) \veq{\epsilon_{ij}(r')}{\epsilon^o_{mn}(r)} d^3r' 
	&= \int (\tilde{\sigma}_{ij} G_{ijkl} C^o_{klmn}) e^{-i\vec{k}.\vec{r}} \din{k}
\end{align}
The integrand $(\tilde{\sigma}_{ij} G_{ijkl} C^o_{klmn})$ of the RHS can be further simplified by substituting for stress as follows:
By taking the Fourier transform of the total strain field $\epsilon_{ij}(\vec{r})$,
 we get,
\begin{align}
\label{deq1}
\tilde{\epsilon}_{ij} &= \bar{\epsilon}_{ij} V \delta_{\vec{0},\vec{k}} + H_{ijkl} C^o_{klmn} \tilde{\epsilon}_{mn}^o
\end{align}
where  $V$ is the Volume of the system,
 $\tilde{\epsilon}_{ij}$ and $\tilde{\epsilon}^o_{mn}$ are the Fourier transforms of the fields ${\epsilon}_{ij}$ and  ${\epsilon}^o_{mn}$, respectively. 
 The kronecker-delta function $\delta_{\vec{k}_1,\vec{k}_2}$ is defined as $1$ if $\vec{k}_1=\vec{k}_2$ and zero otherwise.
From the Hook's law, we know
\begin{align}
	\tilde{\sigma}_{ij}=C^o_{ijop} (\tilde{\epsilon}_{op}-\epsilon_{op}^{oo} \tilde{\theta})
\end{align}
Substituting the Eq.\eqref{deq1} into the above equation, we get
\begin{align}
	\label{deq2}
	\tilde{\sigma}_{ij}=C^o_{ijop} \bar{\epsilon}_{op} V \delta_{\vec{0},\vec{k}} + C^o_{ijop} H_{opkl} C^o_{klmn} \tilde{\epsilon}^o_{mn} - C^o_{ijop} {\epsilon}_{op}^{oo} \tilde{\theta}.
\end{align}
Simplifying the integrand $(\tilde{\sigma}_{ij} G_{ijkl} C^o_{klmn})$ of Eq.\eqref{deq8} using Eq.\eqref{deq2}, we get
\begin{align}
\nonumber
\label{deq3}
\tilde{\sigma}_{ij}G_{ijkl}C^o_{klmn} =G_{ijkl} C^o_{klmn} C^o_{ijop} \bar{\epsilon}_{op} V \delta_{\vec{0},\vec{k}} 
&+ G_{ijkl} C^o_{klmn} C^o_{ijop} G_{opqr} C^o_{qrst} \tilde{\epsilon}_{st}^o \\
&- G_{ijkl} C^o_{klmn} C^o_{ijop} \tilde{\epsilon}_{op}^o 
\end{align}
The first term is identically zero as $G_{ijkl}=0$ when $\vec{k}=0$ and $\delta_{\vec{0},\vec{k}}=0$ when $k \neq 0$. We can then simplify the second term using
\begin{align*}
	G_{ijkl}C^o_{ijop}G_{opqr}&=n_i \Omega_{jk} n_l C^o_{ijop} n_o \Omega_{pq} n_r \\
	&=\underbrace{(n_i C^o_{ijop} n_o)}_{\Omega^{-1}_{jp}} \Omega_{jk} n_l \Omega_{pq} n_r \\
	&= (\Omega_{jk} \Omega^{-1}_{jp}) n_l \Omega_{pq} n_r \\
	&= \delta_{kp} n_l \Omega_{pq} n_r \\
	& \delta_{kp} n_l \Omega_{pq} n_r \\
	&= G_{lkqr}.
\end{align*}
Thus Eq.\eqref{deq3} becomes
\begin{align*}
	\widetilde{\epsilon_{ij}}=\bar{\epsilon_{ij}}V\delta_{\vec{0},\vec{k}}+H_{ijkl}C_{klmn}^o\widetilde{\epsilon_{mn}^o}.		(26).\\
\end{align*}
From Eq.\eqref{deq8}, we can observe that 
$$ 	\int \sigma_{ij}(\vec{r'}) \veq{\epsilon_{ij}(r')}{\epsilon^o_{mn}(r)} d^3r' = 0. $$
Substituting the above equation into Eq.\eqref{deq4}, we get
 \begin{align}
\label{kkeyprop}
 \int \sigma_{ij}(\vec{r'}) \veq{\epsilon_{ij}(r')}{\theta(r)} d^3r' = 0.
 \end{align}
This is the key property of the mechanical equilibrium which greatly simplifies our calculation.
Hence, Eq.\eqref{deq5} reduces to 
\begin{align}
   \label{deq9}
   \int \veq{E^{el}}{\epsilon_{ij}^{el}(\vec{r'})} \veq{\epsilon_{ij}^{el}(\vec{r'})}{\theta(\vec{r})} d^3r'
   =  - \sigma_{ij} \epsilon_{ij}^{oo}.
\end{align}
Combining Eqs.(\ref{varEl2}),(\ref{deq6}) and (\ref{deq9}), we obtain the following expression for the variational derivative of the elastic energy
	\begin{align}
	\label{varEl_final}
	\veq{E^{el}}{\theta(\vec{r})}= - \frac{1}{2} \deq{ \Delta C_{ijkl} }{\theta} \epsilon_{ij}^{el}(\vec{r}) \epsilon_{kl}^{el}(\vec{r}) 
	- \sigma_{ij} \epsilon_{ij}^{oo}
	\end{align}

The above equation is identical to the VDEE derived from the LLJ model.
%
%\end{paragraph}

	\section{Discussion}
In order to quantitatively compare the two models (LLJ and WJK), we numerically computed VDEE and used it as the ground truth. 
A two-dimensional (2D) model system with a square precipitate in the center of our computation cell (Fig.\ref{sch1}a) was chosen for the numerical simulations in this study.
The VDEE was evaluated on this system numerically (Section~\ref{method_numerical}) and analytically  (Section~\ref{anal_expressions}).
The system size is $512\times 512$ and periodic boundary conditions were applied in all directions.
%
%	\end{paragraph}

	    \subsection{Numerical method}
\label{method_numerical}
%\begin{paragraph}{ }
\longcomment{
A numerical method for evaluating the variational derivative is used as an independent evaluation of  Leo's \cite{leo1998} and Zhou's \cite{zhou2008} analytical approaches. 
}%end of comment
The numerical method is based on a first-order Euler method. 
The algorithm to evaluate the variational derivative used at a given reference point $\vec{r}_\text{ref}$ for a given $\theta(\vec{r})$ profile is as follows.
\begin{enumerate}
	\item Solve the mechanical equilibrium $\nabla \cdot \boldsymbol{\sigma}=0$ using the iterative equation described in Section~\ref{yao_shen_sec}.
	\item Compute the elastic energy $E^\text{el}=\int \frac{1}{2} C_{ijkl} \epsilon_{ij}^\text{el} \epsilon_{kl}^\text{el} dV$
	\item Change the order parameter at the reference point $\vec{r}_\text{ref}$ by a small variation, i.e., $\theta(\vec{r}_\text{ref}) \Rightarrow \theta(\vec{r}_\text{ref}) + \Delta \theta $.
	\item Repeat Steps (1-3) for different $\Delta \theta$. In this study, we used $\Delta \theta=h,2h,-h,-2h$ where $h=10^{-6}$.
	\item The slope of a linear fit to the $E^\text{el}$ vs $\Delta \theta$ plot is the required variational derivative $\veq{E^\text{el}}{\theta(\vec{r}_\text{ref})}$. 
\end{enumerate}
As the method is based on first-order Euler, smaller values of $\Delta \theta$ is preferred.
It should also be noted that $\Delta \theta$ should not be very small, which can induce truncation errors. 
The value $R^2$ (commonly known as "goodness of fit" in statistical analysis) was used as a measure to quantify the residual error.
As shown in Fig.\ref{hfix_plot}, there is a wide range of values of $h$ for which $R^2=1$.
We used $h=10^{-6}$ to perform our calculation. 
The procedure was first benchmarked using an elastically homogeneous system where we know the exact analytical value.
%\end{paragraph}
 
\pngimageb{hfix_plot}{For a wide range of step sizes ($h=10^{-10}-10^{-3}$), goodness of fit $R^2=1$. For our calculation, we chose $h=10^{-6}$ (marked by the dashed line). }

%\begin{paragraph}{ }
%
The homogeneous system considered is a single crystal having a cubic symmetry with Zener ratio $A=3$, bulk modulus $B=1083.33$ and  shear modulus $G=500$. A stress-free boundary condition was applied.
All the values mentioned in this article are non-dimensional.
Figure~\ref{sch1}(b) shows the change in the elastic energy $\Delta E^{el}$ for different values of $\Delta \theta$ at the reference point $\vec{r}_{ref}=(135,257)$ marked by the red cross in Fig.\ref{sch1}(a).
%8
The slope of the linear fit is $4.697777\times 10^{-2}$ while the expected value from the analytical expression Eq.\eqref{var_der_homo} at this point is $4.697809 \times 10^{-2}$.
Thus, we conclude that the numerical method is sufficiently accurate.
%\end{paragraph}

\pngimagea{1}{0.5}{1.96}{2.12}{(a)Microstructure used in this study. The red cross marks the reference point used in (b). (b) Change in elastic energy $\Delta E^\text{el}=E^\text{el}-E^\text{el}_\text{initial}$ as function of $\Delta \theta$.}

	    \subsection{Evaluation of various analytical expressions against numerical solutions}

%\begin{paragraph}{ }
Now we consider the same 2D microstructure as that shown in Fig.\ref{sch1}(a) in a elastically inhomogeneous system
where the modulus of the precipitate phase is assumed to be smaller than that of the matrix (e.g., $C^\text{ppt}=0.9C^\text{mat}$).
The non-dimensional values used in the simulations are: 
(i) for the matrix phase: Zener anisotropy ratio $A=3$, bulk modulus $B=1083.33$, and shear modulus $G=500$; (ii) for the precipitate phase: $A=3$, $B=975$, $G=450$; (iii) misfit strain $\epsilon^{oo}_{ij}=0.01 \delta_{ij}$.
We computed the VDEE using the analytical expressions given in Section~\ref{anal_expressions} and also with the numerical method described in Section~\ref{method_numerical}.
The results are summarized in Fig.\ref{mu_elas}.
%
%\end{paragraph}

\pngimageb[0.8]{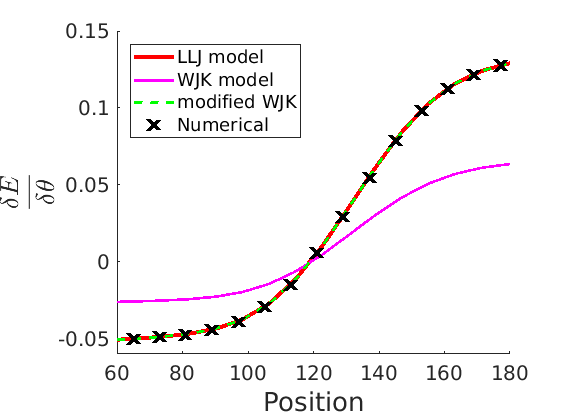}{Comparison among various analytical expressions of the VDEE (variational derivative of elastic energy), $\delta E/\delta \theta$, against the numerical solution. The variational derivatives as a function of position (X) across the red line in Fig.\ref{sch1}(b) for different models: (i) LLJ model Eq.\eqref{veq_leo} \cite{leo1998}, (ii) WJK model Eq.\eqref{veq_zhou} \cite{wang2003phase} (iii) modified WJK Eq.\eqref{veq_Czhou} and (iv) Numerical method (Section~\ref{method_numerical}).}

%\begin{paragraph}{ }
The LLJ model agrees very well with the VDEE calculated from the numerical method as expected.
WJK model has a large deviation from the LLJ model and the numerical method.
However, after adding the first order correction in the virtual strain to the VDEE ("modified WJK"), it agrees well with the numerical value.
%\end{paragraph}
%\begin{paragraph}{ }
By comparing the different expressions of VDEE in  Section~\ref{anal_expressions}, we can see that all the expressions  can be divided into two terms: contribution due to the inhomogeneous modulus ($1^\text{st}$ term) and contribution due to the transformation strain ($2^\text{nd}$ term).
By comparing WJK and "modified WJK", we can see that only the contribution due to the transformation strain is different.
In the WJK model, the virtual strain is treated as a constant within a single time step, which in turn affects the transformation strain contribution to the variational derivative.
The difference between the WJK and "modified WJK" curves in Fig.\ref{mu_elas} arises because of this reason.
%\end{paragraph}
%\begin{paragraph}{ }
%
%
It is interesting to note that even with stronger assumptions in the LLJ model, i.e., neglecting the infinitesimal changes in the total strain due to infinitesimal changes in the order parameter while evaluating the VDEE ,the variational derivative calculation gives the correct result. 
Even though in the LLJ model \cite{leo1998}, the mechanical equilibrium constraint was not imposed while deriving the VDEE, the property we derived in Eq.\eqref{kkeyprop} ensured that the variational derivative given by Eq.\eqref{veq_leo} is identical to the one we derived (Eq.\eqref{varEl_final}).
The proof of  Eq.\eqref{kkeyprop} is based on the Green's operator property $G_{ijkl}C^o_{ijop}G_{opqr}= G_{lkqr}$.
As the Green's operator only has the symmetries of the underlying differential equation (which is $\nabla \cdot \boldsymbol{\sigma}=0$), such a property may hold true even for finite strain calculations.
The property of Eq.\eqref{kkeyprop} makes the assumption in the LLJ model, that the total strain field and the order parameter are independent in calculating variational derivative, unnecessary.
%
%\end{paragraph}	

\subsection{Principle of virtual work}

The property Eq.\eqref{kkeyprop} can be understood from the "principle of virtual displacement", also known as "principle of virutal work" \cite{dym1973solid}. 
The integral on the LHS of Eq.\eqref{kkeyprop} can be correlated with the internal virtual work, which under static equilibrium in the absence of body forces, can be written as,
\begin{align}
\label{virtual_displacement}
    \int\limits_{V} \sigma_{ij} \delta\epsilon_{ij} dV = \int\limits_{S} T_i \delta u_i dS
\end{align}
where $\delta \epsilon_{ij}(\vec{r})$ and $\delta u_i(\vec{r})$ are kinematically admissible virtual strain and displacement fields, respectively,
 and $T_i$ is the traction vector on the surface.
 Note that the virtual strain mentioned here is different from that discussed earlier in Section~\ref{yao_shen_sec}.
 In the context of the principle of virtual displacements, the virtual strain field is the imaginary variation of the total strain field that arises from the infinitesimal perturbations of the displacement field in the real system.
$V$ and $S$ are volume and surface area of the system. 

\pngimageb[0.5]{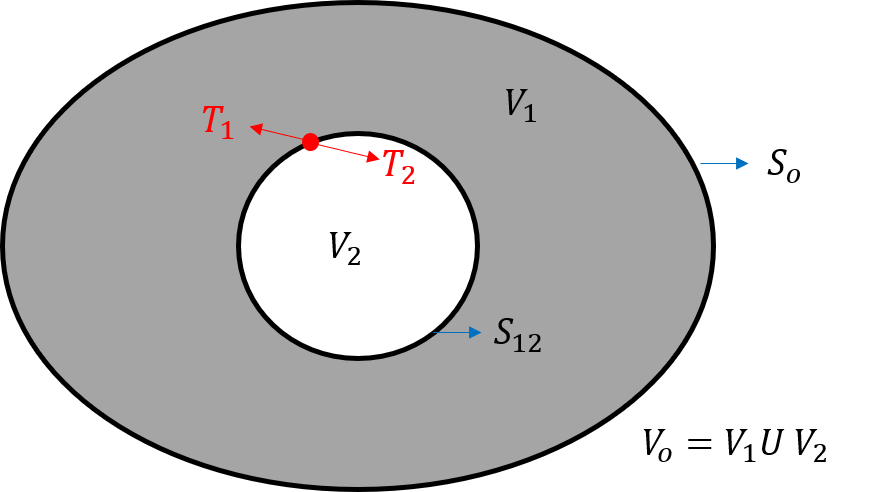}{Schematics of the Eshelby inclusion problem. $V_2$ is the inclusion, $V_1$ is the matrix, $\bar{T}_1$ is the traction vector acting on $V_1$ and $\bar{T}_2$ is the traction vector acting on $V_2$.}

As there can be stress discontinuity on the surface of the  inclusion, the principle of virtual displacements cannot be applied directly to Eshelby's inclusion problem.
However, if we divide the total system into matrix and inclusion, this principle can be applied to each subsystem individually.
For simplicity, let us consider a single inclusion embedded into a matrix as illustrated in Fig.~\ref{eshelby_inclusion}) .

Note that the following arguments can be extended to any number of inclusions. 
By applying the Eq.\eqref{virtual_displacement} individually to the matrix and the inclusion, we get
\begin{align}
\label{body1}
    \int\limits_{V_1} \sigma_{ij} \delta\epsilon_{ij} dV = \int\limits_{S_{12}} (T_{1i}) \delta u_{1i} dS \\
\label{body2}
    \int\limits_{V_2} \sigma_{ij} \delta\epsilon_{ij} dV = \int\limits_{S_{12}} (T_{2i}) \delta u_{2i} dS
\end{align}
where $T_{1i}(\vec{r})$ and $T_{2i}(\vec{r})$ are the "i"-th component of the surface traction on $V_1$ and $V_2$ respectively, and $\delta u_{1i}(\vec{r})$ and $\delta u_{2i}(\vec{r})$ are the "i"-th component of the virtual displacement on the surfaces of $V_1$ and $V_2$, respectively. 
Here we have considered the case where the outer surface $S_\text{o}$ is at infinity and assumed to be stress-free.  We can add the above two equations (Eq.\eqref{body1} and Eq.\eqref{body2}) and, by using Newton's third law ($T_{1i}=-T_{2i}$), we get
\begin{align}
    \int\limits_{V_o} \sigma_{ij} \delta\epsilon_{ij} dV = \int\limits_{S_{12}} T_{1i} [\delta u_{1i}(\vec{r})-\delta u_{2i}(\vec{r})] dS.
\end{align}
We expect the virtual displacement field $\delta u_i(\vec{r})$ to be continuous as it should be kinematically admissible, i.e., $\delta u_{1i}(\vec{r})=\delta u_{2i}(\vec{r})$.
Hence, the RHS of the above equation is zero and we get
\begin{align}
\label{VP_virtual_work}
    \int\limits_{V_\text{o}} \sigma_{ij} \delta\epsilon_{ij} dV=0
\end{align}
In PFM, the total strain field $\epsilon_{ij}(\vec{r})$ is a functional of the order parameter field. By assuming that the variation in the strain field, $\delta \epsilon_{ij}(\vec{r})$, is caused by the perturbation in the order parameter field, $\delta \theta(\vec{r})$, we can write
\begin{align}
\label{VP_vepsilon}
    \delta \epsilon_{ij}(\vec{r})=\int \left(\veq{\epsilon_{ij}(\vec{r})}{\theta(\vec{r'})}\right) \delta \theta(\vec{r'}) d^3\vec{r'}
\end{align}
By substituting the above equation in Eq.\eqref{VP_virtual_work} and simplifying the integral, we get
\begin{align}
    \Huge\int \left(  \left[ \int \sigma_{ij} (\vec{r}) \left(\veq{\epsilon_{ij}(\vec{r})}{\theta(\vec{r'})}\right)  d^3r 
    \right] \delta \theta(\vec{r'}) \right) d^3\vec{r'}
    = 0
\end{align}
For the above equation to hold for arbitrary $\delta\theta(\vec{r})$, the term within the square brackets has to be zero, i.e.,
\begin{align*}
    \int \sigma_{ij} (\vec{r}) \left(\veq{\epsilon_{ij}(\vec{r})}{\theta(\vec{r'})}\right)  d^3r =0.
\end{align*}
Thus we again recovered Eq.\eqref{kkeyprop} through the principal of virtual displacement.
Therefore, the physical foundation for the key property shown by Eq.\eqref{kkeyprop} is the "principle of virtual displacement".

	\section{Conclusion}
%\begin{paragraph}{ }
We have derived a physically transparent and numerically accurate expression of the variational derivative of the elastic energy VDEE in an elastically anisotropic and inhomogeneous media, in which the mechanical equilibrium is used as a constraint.
The new expression as well as the existing ones in the literature are compared and evaluated against numerical solutions. We find that:
\begin{itemize}
    \item LLJ model \cite{leo1998} gives the accurate mathematical expression of VDEE despite its nonphysical assumption, i.e., the total strain field  \Totalstrain is not a function of order parameter \OrderParameter. It is due to the key property Eq.\eqref{kkeyprop}.

    \item WJK model shows a significant deviation from the numerical VDEE due to the constant virtual strain assumption. 
    Thus WJK model may not give the accurate evolution of the PFM order parameters. 
    However, the "modified WJK" model we derived (Eq.\eqref{veq_Czhou}), has sufficient accuracy to reproduce the VDEE from the numerical method.
    Furthermore, we have proved by using the virtual work principle that the assumption made widely in the existing literature, i.e., the total strain field of an elastically anisotropic and inhomogeneous system is independent  of the microstructure of a two-phase mixture, which is obviously not true, is completely unnecessary.

\end{itemize}

%\end{paragraph}

\section*{Data availability statement}
The raw/processed data required to reproduce these findings cannot be shared at this time due to technical or time limitations.
\nocite{kadirvel2022exploration,koneru2022high,kloenne2021high,sriram2023formation,sriram2023formation,lu2021microstructure}
\section*{Acknowledgement}
The authors would like to thankfully acknowledge the financial support from Air Force Office of Scientific Research (AFOSR) under grant FA9550-20-1-0015.

\bibliography{inhom_solver} 
\bibliographystyle{elsarticle-num}

\appendix
		\section{Notations}
	\label{APPnotation}
		\begin{tabular}{cl}
		$C^o_{ijkl}$ &- Reference modulus	\\
		$C_{ijkl}(\vec{r})$ &- Inhomogeneous modulus \\
		$S_{ijkl}(\vec{r})$ &- Inhomogeneous compliance \\
		$\sigma_{ij}(\vec{r})$ &- Local stress \\
		$\epsilon_{ij}(\vec{r}$) &- Total strain \\
		$\epsilon_{ij}^{oo}$ &- Eigenstrain \\
		$\epsilon_{ij}^T(\vec{r})=\epsilon_{ij}^{oo} \theta(\vec{r})$ &- Transformation strain \\
		$\sigma_{ij}^T(\vec{r})=C^o_{ijkl} \epsilon_{ij}^T$ & Transformation stress \\
		$\epsilon_{ij}^{el}(\vec{r})$ &- Elastic strain \\
		$\epsilon^o_{ij}(\vec{r})$ &- Virtual strain \\
		$\sigma_{ij}^o(\vec{r})=C^o_{ijkl} \epsilon_{kl}^o$ &- Virtual stress \\
		$\Delta C_{ijkl}(\vec{r})=C^o_{ijkl}-C_{ijkl}(\vec{r})$ & \\
		$\Delta S_{ijkl}(\vec{r})=(\Delta C_{ijkl})^{-1} $ & \\
		\end{tabular}
	%\add{a2_keyprop}
	%\add{a3_questions}
		
\end{document}